# *Impact of nitrogen seeding on confinement and power load control of high triangularity JET ELMy H-mode plasma with a metal wall*

C. Giroud[1], G.P. Maddison[1], S. Jachmich[2], F. Rimini[1], M. Beurskens[1], I. Balboa[1], S. Brezinsek[3], R. Coelho[4], J.W.Coenen[3], L. Frassinetti[5], E. Joffrin[6], M. Oberkofler[7], M. Lehnen[3], Y. Liu[8], S. Marsen[9], K. McCormick[7], A. Meigs[1], R. Neu[7], B. Sieglin[7], G. van Rooij[10], G. Arnoux[1], P. Belo[4], M. Brix[1], M. Clever[3], I. Coffey[11], S. Devaux[7], D. Douai[6], T. Eich[7], J. Flanagan[1], S. Grünhagen[1], A. Huber[3], M. Kempenaars[1], U. Kruezi[3], K. Lawson[1], P. Lomas[1], C. Lowry[12], I. Nunes[4], A. Sirinnelli[1], A.C.C. Sips[11], M. Stamp[1], S. Wiesen[3] and JET-EFDA contributors*

*JET-EFDA, Culham Science Centre, Abingdon, OX14 3DB, UK.*
*[1] EURATOM/CCFE Fusion Association, Culham Science Centre, Abingdon, Oxon. OX14 3DB, UK.*
*[2]Association Euratom-Etat Belge, ERM-KMS, Brussels Belgium.*
*[3]IEK-Plasmaphysik, Forschungszentrum Jülich, Association EURATOM-FZJ, Jülich, Germany.*
*[4]IPFN, EURATOM-IST Associação, 1096 Lisbon, Portugal.*
*[5]Association EURATOM-VR, Department of Physics, SCI, KTH, SE-10691 Stockholm, Sweden.*
*[6]CEA-Cadarache, Association Euratom-CEA, 13108 St Paul-lez-Durance France.*
*[7]Max-Planck-Institut für Plasmaphysik, EURATOM-Association, 85748 Garching, Germany.*
*[8]Institute of Plasma Physics, Chinese Academy of Sciences, Hefei 230031, China.*
*[9]Max-Planck-Institut für Plasmaphysik, Teilinsitut Greifswald, EURATOM-Assoziation, D-17491 Greifswald, Germany*
*[10]Association EURATOM-FOM, PO BOX 1207, 3430 BE Nieuwegein, The Netherlands.*
*[11]Astrophysics Research Centre, Queen's University, Belfast, BT7 1NN, Northern Ireland, UK.*
*[12]JET-EFDA/CSU, Culham Science Centre, Abingdon, OX14 3DB, U.K.*
*[*]see Appendix of F Romanelli et al, Fusion Energy (24th IAEA Fusion Energy Conference 2012, San Diego, US*
*Email:carine.giroud@ccfe.ac.uk*

*Abstract*: *This paper reports the impact on confinement and power load of the high-shape 2.5MA ELMy H-mode scenario at JET of a change from an all carbon plasma facing components to an all metal wall. In preparation to this change, systematic studies of power load reduction and impact on confinement as a result of fuelling in combination with nitrogen seeding were carried out in JET-C and are compared to their counterpart in JET with a metallic wall. An unexpected and significant change is reported on the decrease of the pedestal confinement but is partially recovered with the injection of nitrogen.*

## 1. Introduction

The reference scenario for achieving Q=10 in ITER is the type-I ELMy H-mode at 15MA, with a deuterium-tritium mixture, sufficient energy normalised confinement ($H_{98(y,2)}$), at high density ($H_{98}$~1.0, Greenwald fraction $f_{GDL}$~0.85) and compatible with its Plasma Facing Components (PFCs). The expected power to the divertor makes high divertor radiation mandatory and requires the injection of extrinsic impurities. The challenge of achieving the scenario requirements with the material selection of the DT phase of ITER is being addressed at JET with its new wall (JET-ILW) including main-chamber limiters made of bulk beryllium (Be) and a full tungsten (W) divertor [1,2]. Three aspects require investigation: first the capacity to reduce the inter-ELM power load to the W divertor without significant degradation of the energy confinement; second the limitation of impurity production from the new PFCs to maintain performance, in terms of dilution (Be) and core radiation (W), the latter by reducing the energy of impurity ions in divertor which governs the W sputtering yield [3]; third the compatibility of the extrinsic impurities with the ILW materials in particular in the case of a species such as $N_2$.

The choice of an extrinsic impurity for ITER is not only motivated by its impact on the plasma confinement with a high divertor radiation but also by its interaction with the PFCs and its consequences for the tritium processing plant. Being chemically reactive, $N_2$ can interact with the PFCs to form nitrides but the more important issue is its chemical reaction with hydrogen isotopes to form ammonia ($ND_3$) [4]. The produced amount of ammonia is potentially considerable and would have implications for the design of the ITER tritium plant. The ITER-team plans to have the flexibility to use Ne, Ar and $N_2$ (or a mixture of) as seed gases, but it remains the case that chemically reactive species will be more of a safety challenge for the plant than a noble gas. Nevertheless, nitrogen as extrinsic radiator has a role to play in the development of radiative scenario in present devices for several reasons: it plays the role that Neon will have in ITER divertor conditions where the pedestal will be hotter ~3-4 keV [5]; it also isolates the effect that high divertor radiation for power load reduction has on an integrated scenario. At JET, Ne and Ar also radiate in the pedestal and main plasma (respectively) adding an additional degree of difficulty in achieving stationary conditions due to their ability to reduce the power entering the edge region and as a result reduce the pedestal confinement [6][7].

This paper investigates the potential of $N_2$ seeding for achieving a radiative divertor for power load control at JET with the new metal wall. It focuses on the ELMy H-mode scenario at high triangularity (2.7T/2.5MA, $q_{95}$~3.5, $P_{in}$~16MW, $\delta$~0.4). The high triangularity scenario was chosen as it had the following characteristics in JET-C: its global confinement enhancement factor $H_{98(y,2)}$ is higher than the low triangularity scenario at the required Greenwald fraction for ITER of ($f_{GLD}$~0.85); it can be fuelled up to Greenwald density ($f_{GDL}$~1.0) without confinement loss ($H_{98}$~1.0)) at JET, the so-called type-I/II ELM regime [8].

This paper is organized as follows: a summary of the seeded experiments in JET with Carbon-Fibre Composite PFCs (JET-C) is given in section 2; differences in confinement between JET-C and JET-ILW are reported in section 3; the recovery of confinement with $N_2$-seeding in presented in section 4; the inter-ELM power load reduction resulting from $N_2$ seeding is shown in section 5; the lack of stationarity of the plasma discharges and the averaged erosion yield is discussed in section 6; the lack of improved confinement with $N_2$ seeding in low triangularity plasma is briefly discussed in section 7 together with the possible effect of dilution in the confinement improvement; a concluding summary is given in section 8.

## 2. Description of experiments and main results from JET-C seeding studies

In preparing to document the change of wall and its impact on power load to the divertor, experiments in JET-C explored the reduction of the inter-ELM power load in an ELMy H-mode scenario (2.7T/2.5MA, $q_{95}$~3.5, $P_{in}$~16MW, $\delta$~0.4, $H_{98(y,2)}$~1.0, $f_{rad}$~0.45) at high density ($f_{GDL}$~0.8) with a fuelling mix of $D_2$ and either Ne or $N_2$ as seed impurity, constant input power and proximity to the transition from type I to III ELMs [6,9], see Fig. 1. A deuterium fuelling scan from 0.3 to 6.0x$10^{22}$ el/s was performed during part of which the ELM regime changed from type-I, with a frequency of 20-25Hz (at lowest fuelling level), to the so-called type I/II with ELM frequency of 7-10Hz [6].

It was shown that with either $N_2$ or Ne, conditions of partial divertor detachment were achieved with less than 10% energy confinement reduction in the main plasma with respect to low-fuelled reference case (~0.3x$10^{22}$el/s, #76666). A difference in the radiation localization was observed between the two seeding gases. $N_2$ seeding increased the divertor radiation while Ne seeding decreased it. Although reduced with the injection of extrinsic impurity, C concentration remained above 1% and contributed to the divertor radiation in addition to the extrinsic impurity[6]. Also it was observed that the H-mode pedestal at the plasma edge of Ne-seeded discharges often crossed the boundary between type-I to III ELM regime, leading to compound ELMs and unsteady edge conditions. No condition was observed in this configuration for which $N_2$-seeding increased the plasma confinement, as was observed in ASDEX Upgrade with a full W-wall [10].

The transition from type-I to type-III ELM regime turned out to be the limitation in achieving a high value of divertor over main plasma radiation with an H-factor close to 1.0. The main plasma, $P^{bulk}_{rad}$, and divertor, $P^{div}_{rad}$, radiated power were determined from tomographic reconstruction by integrating the 2D reconstruction for z-abscissa value greater than -1.2m for $P^{bulk}_{rad}$ and for $P^{div}_{rad}$ with z less than -1.2m, as described in [6]. Fig. 1 shows that when $N_2$ is seeded in JET-C, at a rate from 0.0 to 2.4x$10^{22}$ el/s, in the two plasma with highest density (#76684 and #76678), the pedestal density decreases followed by a decrease of both temperature and density at the highest seeding rate of ~4x$10^{22}$ el/s for which a transition from type-I to type-III ELM regime occurred [6].

## 3. Differences in confinement between JET-C and JET-ILW high-shape ELMy H-mode discharges at 2.5MA

Similar discharges were repeated in JET-ILW with practically identical high-shape (2.5MA/2.7T) with the aim to characterize the difference between JET-C and JET-ILW. Studies have shown that C concentration has dropped in the divertor and main plasma by a factor 10 [11]. Although dedicated studies address specific issue related to a change of material (such as divertor detachment [12], W erosion[13,14], impurity composition[15]), the experiments presented here integrate all the aspects and provide a comparison with JET-C of the overall impact of the new PFCs on the chosen plasma H-mode scenario. As expected, W contamination in the main plasma makes operation at low fuelling difficult and high-shape ELMy H-mode at 2.5MA is restricted to fuelling levels higher than $0.9\times10^{22}$ el/s in contrast to the lowest fuelling of ~$0.3\times10^{22}$el/s accessible in JET-C related studies. In any case, studies made in JET-C in preparation for the change of PFCs provides a set of reference discharges for higher fuelling levels (up to $6\times10^{22}$el/s).

A comparison of two plasmas with similar starting parameters between JET-C (#76678) and JET-ILW (#82806) plasma at 2.5MA at a $D_2$-fuelling rate of ~$2.7\times10^{22}$ el/s is shown in Fig. 2. At similar average electron density, and similar input power, the stored energy of the JET-ILW discharge is reduced by 40% in comparison to JET-C with a stored energy of 3.8MJ instead of 6.3MJ. The total radiative power has clearly decreased and the radiative fraction has dropped from $P_{rad}/P_{in}$=0.45 to 0.30. The radiative power from the main and divertor plasmas is reduced in comparison to the JET-C reference discharge, by 1.4MW and 2.1 MW respectively. In JET-ILW, the plasma is cleaner with the effective charge $Z_{eff}$ reduced from 1.7 to 1.3 owing to the main impurity being now Be instead of C. The neutral pressure in the sub-divertor is similar in both discharges. The drop in confinement in the JET-ILW plasma compared to that for JET-C cannot be attributed to an increased main plasma radiation due to W. In fact, it stems from a drop in pedestal pressure and mostly temperature. As shown in Fig. 1, the pedestal electron temperature has dropped from a value of 0.6keV down to 0.4keV. More generally, similar plasmas within a fuelling range of 0.9 to $3.2\times10^{22}$ el/s in the ILW covers a similar range of pedestal electron density as in JET-C, as shown in Fig. 1, but the pedestal electron temperature is on average 30-40% lower. Although a reduction of the operational window was foreseen if main plasma radiation increased due to W contamination, this impact of the ILW on the confinement in these high triangularity plasmas was not anticipated.

Even more surprising is the fact that JET-ILW plasmas with type-I ELMs, such as #82806, exists below the critical electron temperature for type-III ELMs in JET-C at high density, see Fig. 1 and 3 [6]. These plasmas have benign ELMs with a slower drop in pedestal pressure than anything noted before [1] [16] and seem to exist in the pedestal $n_e$ and $T_e$ space which in JET-C would have been occupied by the type-III ELM regime. The ELM stability regime seems to have change significantly. The slow ELMs are discussed in more details in [1] and [16]. The boundary between the type-I to type-III ELM regime which was such a key player in both the ELM dynamics and drop in confinement in JET-C discharges as a result of seeding is not yet fully characterized for JET-ILW, but lies far below its previous position.

## 4. Recovery of confinement with $N_2$ seeding in high-shape ELMy H-mode discharges

Although nitrogen was first injected to increase the divertor radiation and to reduce the power loads, nitrogen seeding has been found to improve plasma energy confinement in high-shape ELMy H-mode. Fig. 4 shows time traces of comparable plasmas, seeded with $N_2$ in JET-ILW and an unseeded JET-C reference case at similar pedestal density. When nitrogen is injected into deuterium-fuelled discharges in JET-ILW, it raises the pedestal density and temperature leading to an increase in stored energy to 5.5MJ close to the JET-C $D_2$-fuelled counterpart, i.e. #76684 with 5.8MJ. The ELM frequency drops from ~16Hz to 4Hz is a result of seeding (was 9Hz for #76684). Without seeding, ELM crashes are slow and lead to benign power load on the divertor. With $N_2$-seeding, however, the ELM crashes speed up again and make the ELMs resemble more those seen with the carbon wall [16]. For the range of fuelling level of 0.8-3.0x$10^{22}$ el/s, it can be seen in Fig. 5 that nitrogen seeding allows access to higher electron pedestal temperatures, only slightly lower than deuterium-fuelled JET-C counterparts. This leads to stored energies and H-factors only slightly below their deuterium-fuelled counterparts as seen in Fig. 6 and 7, and a very good match to JET-C $N_2$-seeded pulses at similar nitrogen seeding rates, particularly when the density dependence is removed from the $H_{98}$-factor scaling to allow the energy confinement time to be compared more directly to plasmas of different density, as shown in Fig. 7. The electron density of seeded discharges tends to be higher in JET-ILW than JET-C for same seeding level. The best $N_2$-seeded pulse with the JET-ILW gives $H_{98}$~0.9, $n/n_{GW}$~1.0 with $Z_{eff}$~1.5. The plasma stationarity is not achieved however and is addressed in section 7.

This is the first time in JET that injection of impurities leads to an increase in global energy confinement. Such a strong correlation between nitrogen seeding rate and stored energy has already been observed in ASDEX-Upgrade and is reported there to be linked to a positive correlation between H-factor and $Z_{eff}$ [10]. In JET-ILW, such a clear dependence with $Z_{eff}$ has not been demonstrated and is discussed in more details in section 7.

It is important to note that the increase in confinement as a result of the nitrogen injection in the ILW stems from the pedestal and not from an improved energy and particle confinement in the core plasma. The peaking of the electron density and temperature have been calculated taking the ratio of the averaged quantity at r/a ~0.4 over its value at r/a~0.8 and is shown in Fig. 8. No change in the electron density peaking is observed between the fuelled only discharges in the JET-C and JET-ILW and the seeded discharges in the JET-ILW. The electron temperature peaking for the fuelled JET-C discharges varies but cover similar range to the JET-ILW discharges whether fuelled only or with the addition of seeding. This indicates that the confinement in these plasmas is set by the achieved pedestal pressure in combination with profile stiffness.

The JET-ILW results strongly suggest that the carbon impurities played a role in the performance of the high-shape plasma scenario in the JET-C era. This is all the more striking if a closer look is taken at the plasma trajectory in the pedestal $T_e$-$n_e$ diagram following $N_2$ injection together with the ratio of divertor to main plasma radiative power (as described in section 2), $P_{rad,div}/P_{rad,bulk}$, in the inter-ELM phase, as shown in Fig. 5 and 10. Focusing on the two highest fuelling levels in the JET-C plasmas, one can see that as $N_2$-seeding rate is increased, the ratio $P_{rad,div}/P_{rad,bulk}$ rises as expected and the pedestal confinement is slightly reduced up to the transition to type-III ELMs regime where the pedestal confinement drops by 20% at the highest divertor radiation. In comparison for JET-ILW plasmas as nitrogen seeding rate is increased, the ratio $P_{rad,div}/P_{rad,bulk}$ is raised and the pedestal confinement increases up to values similar to those in the deuterium fuelled JET-C counterparts. Once the maximum pedestal pressure has been achieved with nitrogen seeding, a further raise in seeding level leads to a decrease in density and a weaker decrease in temperature as shown in Fig. 5, very like the trajectories of plasmas in JET-C with $N_2$-seeding. Similarly the ELM frequency then start to increase in the JET-ILW as was the case for the JET-C plasmas. This is discussed in more detailed in a follow-up publication [17]. For completeness, the effective charge $Z_{eff}$ is shown in Fig. 11 and shows that with increased $N_2$-seeding rate, $Z_{eff}$ is increased in JET-IW but remains at the highest seeding rate lower than the fuelled JET-C counterpart. These observations indicate that C concentration and associated radiation may have been a hidden parameter in the JET-C confinement behaviour in high-shape fuelled discharges, and that the injection of $N_2$ in JET-ILW partially recovers this effect. The exact mechanism is unknown. It remains to be identified whether the effect is a result from the ion dilution, which is addressed in section 7, a change in resistivity or local current density or possible turbulence suppression.

## 5. Inter-ELM power load reduction in JET-ILW high triangularity discharges

Although the increase in pedestal stored energy is a welcome benefit of seeding, nitrogen injection was first envisaged as a way to increase the divertor radiation and thereby reduce the inter-ELM power load to the divertor. With the factor 10 reduction of C in the ILW, the resulting reduced divertor radiation is expected to lead to higher power loads and higher electron temperature at the outer target in the inter-ELM phases.

In JET-C, for the $D_2$-fuelling rate of $2.7x10^{22}$ el/s (#76678), the radiative fraction is ~0.5 (see Fig. 12) with a ratio of divertor to main plasma radiative power ($P_{rad,div}/P_{rad,bulk}$) of ~0.7 in the inter-ELM phase, as shown in Fig. 10. In JET-ILW at the same $D_2$-fuelling rate as above (#82806), the radiative fraction is down to ~0.3 with a ratio ($P_{rad,div}/P_{rad,bulk}$) also reduced down to ~0.3. As the $N_2$-seeding rate is increased in successive discharges in JET-C, the radiative fraction is increased to the highest value of ~0.55 ($N_2$-seeding rate of ~$2.4x10^{22}$ el/s) whilst still in type-I ELM regime with a ratio $P_{rad,div}/P_{rad,bulk}$ of ~0.78. Similarly for the JET-ILW discharges with the same fuelling rate and at nitrogen seeding rate ~2.35 $x10^{22}$ el/s at the highest pedestal stored energy, the radiative fraction reached ~0.52 (#82810) with $P_{rad,div}/P_{rad,bulk}$ of ~0.78, similar conditions in terms of radiative fraction and distribution to the JET-C counterpart. At the higher seeding rate of ~$3.7x10^{22}$ el/s, the $f_{rad}$ reached ~0.54 (#82811) with $P_{rad,div}/P_{rad,bulk}$ of ~0.90 with a somewhat degraded pedestal stored energy. In fact, independently of the fuelling rate, with a high enough nitrogen seeding rate in the JET-ILW, similar conditions of radiative fraction and value of the ratio $P_{rad,div}/P_{rad,bulk}$ can be obtained in JET-ILW as was obtained the JET-C $N_2$-seeded studies, see Fig. 10 and 12. It is then expected that the power load would be most different between JET-C and JET-ILW at no or low $N_2$-seeding and become similar at high level of seeding.

If the focus now is on the measured power landing on the outer target in the inter-ELM phase, we need to study the Infra-Red (IR) or Langmuir Probes (LP)measurements. In the JET-C discharges, the power loads were monitored by IR. The IR measurements were validated against thermocouple data and an energy balance was carried out. It was found that the measured energy loss was 80% of the input energy over the $N_2$ seeding scan [6]. In other words, 20% of the input energy was unaccounted for in the energy balance. For the JET-ILW discharges, technical difficulties linked to the high $\delta$ configuration meant that IR measurements were not available. The thermocouples, embedded in the tile on which outer strike point (OSP) was, were unreliable. The power loads then has to be monitored by Langmuir probes (LP)[18]. The power load in the inter-ELM period can now be compared between the JET-C and JET-ILW discharges. A higher power load is expected for JET-ILW than JET-C for similar fuelling level. The power $P^{inter}_{sep}$ flowing through the separatrix in the inter-ELM phase can be calculated with $P^{inter}_{sep}=P^{inter}_{tot} - P^{inter}_{rad,main}-dW^{inter}_{mhd}/dt$ with $P^{inter}_{tot}$ the total input power (taking into account the shine-through losses but not the fast-ion losses), $P^{inter}_{rad,main}$ the main plasma radiated power and finally $dW^{inter}/dt$ the inter-ELM stored energy build-up rate. Relying on results presented in Ref. [6] for JET-C discharges, at the fuelling level of $D_2$-fuelling of $2.7x10^{22}$ el/s (#76678), the power flowing through the separatrix in the inter-ELM phase, $P^{inter}_{sep}$, is ~8.5MW calculated from a total input power of 15.8MW to which $P^{inter}_{rad,main,}$ of ~4.8MW and $dW^{inter}/dt$, of ~ 2.3MW terms were subtracted. The power landing on the outer target plate is of 1.9MW and therefore a fraction $f^{inter}_{div,OT}$ ~0.22 of $P^{inter}_{sep}$ reaches the outer divertor. The JET-ILW discharge with the same fuelling rate, has a similar power $P^{inter}_{sep,}$ of ~7.6MW calculated with an input power of 16MW, with $P^{inter}_{rad,main}$ ~3.5MW and $dW^{inter}/dt$ ~ 5.6MW, as shown in Fig 13. The power reaching the outer target is 2.4MW and a fraction 0.3 of $P^{inter}_{sep}$. The electron temperature at the outer strike point is ~26eV in JET-ILW compared to ~5eV for its JET-C counterpart [6]. As expected, the power load and electron temperature is higher in JET-ILW than in the JET-C plasmas for a similar fuelling but weaker increase as could have been expected from a reduction by a factor 10 of the C content. Related studies are currently investigating this question [12].

When $N_2$-seeding rate is increased in the JET-C discharges up to $\sim 2x10^{22}$ el/s (focusing here on discharge with type-I ELM regime at $D_2$-fuelling rate of $2.2x10^{22}$ el/s), the power $P^{inter}_{sep,}$ is slightly decreases due to the increase of the $dW^{inter}/dt$ term to 7.2MW. Partial detachment was reached at this seeding level as shown in Ref. [6]. The peak power load is reduced down from $2.8MW.m^{-2}$ (#76678) to below $1MW.m^{-2}$ (#76680) at that $N_2$-seeding rate of $2.4x10^{22}$ el/s and within the error of the IR measurements. For the JET-ILW counterpart with the same fuelling rate, as $N_2$-seeding rate is raised, the power $P^{inter}_{sep}$ is fact increased slightly from 7.6MW to 9.4MW as the $dW^{inter}/dt$ decreases with a reduction of the ELM frequency. The power measured on the outer divertor is of 1.0 MW (#82812) and a decrease from 0.3 to 0.1 of $P^{inter}_{sep}$. With additional seeding of $3.7x10^{22}$ el/s, the ratio $f^{inter}_{div,OT}$ can be reduced down to 0.03 bearing in mind that the inter-ELM phase become difficult to measure for discharge #82811 as the ELM frequency rises again. In any case, the reduction of the power at the outer target is achievable at high enough $N_2$-seeding and a wide $D_2$ fuelling rate of 0.8-2.5x$10^{22}$ el/s.  The electron temperature is reduced to ~7eV (#82812) and reaches similar conditions to those in discharge #76680 with partial detachment and in type-I ELM regime. More details studies on detachment will be presented in a follow-up publication [17].

## 6. Stationarity of plasma and average W erosion yield

The W release is governed by impurity ions due to their threshold energy for W sputtering being an order of magnitude lower than that for deuterium. The sputtering yield for impurity ions such as $C^{4+}$, $Be^{2+}$, $N^{4+}$ rises dramatically for plasma temperature above ~5eV [19,20]. At ~5eV, it is expected that no significant W will be sputtered by low Z impurities in between ELMs. Tungsten will, however, still be sputtered by ELMs.

In these experiments, W erosion was measured by means of passive emission spectroscopy. The WI line radiation at 400.9nm is monitored by a mirror-link system [21] with a time resolution of 40ms, mostly able to provide an ELM-averaged measurement. The measured light intensities were transformed into W particle flux densities using the number of ionization per emitted photons, i.e. S/XB value as in reference [22]. The effective erosion yield is then obtained by normalization of the ELM-averaged peak W particle flux to the saturation current measured by LP. Results are shown in Fig. 14.  It can be seen that for fuelled only plasma (open symbol) where the pedestal pressure is weak and ELM benign, the erosion does decrease with target temperature as would be expected for averaged erosion dominated by the inter-ELM phase. When $N_2$ is seeded, the effective erosion increases even though the electron temperature is decreased to value of ~5eV, thus all sputtering is due to high energetic ions hitting the target plate during ELMs [13]. ELM energy density responsible for material erosion is directly related to the pedestal pressure [23]. In fact the increased erosion yield at low target temperature is directly linked to an increase in pedestal confinement and therefore pressure as shown in Fig. 14. In other words, as nitrogen is seeded and the pedestal confinement increases, the W erosion yield is governed by the ELMs and not the inter-ELM phase; This is compounded by the fact that transport of W into the core through the pedestal region is a critical step in a chain which makes it a dominant factor affecting the build-up of W in normal H-modes [24,25]. Previous studies in $D_2$-fuelled discharges (in which Be is the main impurity responsible for W sputtering) have shown that the intra-ELM W sputtering dominated by a factor 5 over the inter-ELM phase [13]. In our discharges, N is present, and being a heavier than Be, is expected to sputter W more than Be during an ELM.

The $N_2$-seeded high $\delta$ ELMy H-mode considered provides the first case in JET-ILW of a scenario with power load control though high ELM energy density loads together with long enough plasma duration to probe the impact of W on plasma confinement and stationarity. Time traces of a typical $N_2$-seeded discharge with $H_{98}$~0.9, $n/n_{GW}$~1 and $Z_{eff}$~1.5 are shown in Fig. 15. This figure illustrates that $N_2$-seeded high-$\delta$ ELMy H plasmas with higher confinement so far have an unstationary behaviour. Although the total radiative power and energy confinement are constant over most of the plasma duration, the core electron temperature decreases, while the plasma radiation within the sawtooth (ST) inversion radius increases. Once the sawteeth disappears, the core radiation increases exponentially. W is accumulating in the very plasma centre (within rho~0.3) and when the plasma enters its termination phase leads to high radiative power and finally a disruption. It is important to note that high-$\delta$ ELMy H-modes were not stationary in JET-C either with a $Z_{eff}$ constantly increasing as shown in Fig. 2. In fact, ICRH was added to neutral-beam heating as an integral part of the scenario for control of density peaking and sawteeth [8,26]. In the JET-ILW discharge, first test with ICRH heating with a rather limited maximum power of ~2MW (due to technical problems) did not show much benefit, but further investigations will be carried out in the future. The density peaking outside the ST radius is the same in ILW as in JET-C. As a result, it could be that the impurity peaking is similar to what was observed in the JET-C, just made more evident by the presence of an additional and much higher Z impurity such as W.

To develop a long discharge of about 20s flat-top it is essential to improve the plasma stationarity by reducing first W penetration through the separatrix and second W peaking inside the main plasma. The first point needs to be addressed by a reduction of the W source or transport from source to pedestal both of which can be addressed with different plasma divertor configurations. The second point is likely to require the use of core heating for either controlling the ST behaviour or some possible beneficial outward turbulent convection.

## 7. Lack of improved confinement with $N_2$ seeding in other configurations and discussion of possibility for dilution being reason for improvement confinement

Observations presented so far have indicated that C level and associated radiation may have been a hidden parameter in the JET-C confinement behaviour in high-shape fuelled discharges. This could be why such discharges in JET-ILW up to ~40% decrease in confinement compared to the JET-C counterpart, and why the injection of $N_2$ partially recovers this confinement. In contrast, it is not believed that C radiation is a hidden parameter for low-δ ELMy H-mode discharges. Their confinement is reported to be comparable in JET-ILW and JET-C [16], the only difference being that low fuelling operation, favourable for high normalised confinement, is no longer accessible due to W contamination in the main plasma. Nevertheless as a mean to gather further evidence on the effect of nitrogen, this seed impurity was injected into low–δ ELMy H-mode at 2.5MA and similar divertor geometry. A low-δ ELMy H-mode with vertical targets, more similar to the ITER divertor geometry, was also tested. No significant increase in confinement with $N_2$-seeding was observed. The range of $Z_{eff}$ values covered with the low δ configuration was from 1.2 to 1.5 and similar to that of the high δ configuration (1.1-1.8) (see Fig, 11), demonstrating the lack of dependence of confinement on $Z_{eff}$ value.

The vertical target configuration proved interesting for two reasons. First the W contamination in the main plasma was much reduced and the operational window seemed to extend to very low fuelling close to levels used in JET-C. Very first results show signs of a higher stored energy with no additional fuelling but further work is needed to confirm this. Second, W radiation events (attributed to small particle of medium and high Z material entering the plasma [Matthews1]) fairly frequently with the horizontal target divertor configuration [1] but with the vertical target none were observed so far. These results will be presented in details in future publications.

*Discussion on a possible dilution effect to explain increase in stored energy:* In this paper we presented the increase in stored energy following the injection of nitrogen as an effect due to the increase in pedestal stored energy by studying the pedestal electron density and temperature implying tacitly that the main ion temperature and density has a similar behaviour. It is indeed a possibility that the measured increased in the electron channel could not be related to an increased in the total pedestal pressure but to a dilution effect as is advanced for ASDEX-Upgrade to explain the increase in confinement following $N_2$ seeding [10]. For the same of argumentation, if we were to make this hypothesis and state that there is not total improvement of the pedestal pressure we still need to be able to find the conditions compatible with the experiments, meaning the same $Z_{eff}$, $T_{e,ped}$ and $n_{e,ped}$ increase, and total increase in the stored energy. Assuming that the total pedestal pressure $P_{ped,}$ remains unchanged with the injection of nitrogen:

$$P_{ped}= k_B \times (n_{e,ped}\, T_{e,ped}+n_{i,ped}\, Ti_{,ped})= k_B \times T_{ped} \times (n_{e,ped} +n_{i,ped} )=C$$

with $n_{i,ped}$, the total ion density, and $T_{i,ped}$, the ion temperature. Here $T_{e,ped}=T_{i,ped}$ is taken as is confirmed by charge exchange recombination measurements. The ratio $T_e/T_i$ is not measured due to the absence of core ion temperature measurements, but $T_e/T_i$ equal to 1 can also be expected for the whole of the profile since the equipartition time between ion and electron are an order of magnitude lower than the energy confinement time at the high densities considered here. The increase in $T_{e,ped}$ of a factor 1.4 (ignoring the increased of density from now on $n_{e,ped}$ of a factor 1.1) measured in the experiment (#82810 compared with #82806) would need to be compensated by a decrease in $n_{i,ped}$ by about at least 1.4. Assuming nitrogen (with atomic number $z_k$~7) to be the main impurity ions, and invoking quasi-neutrality:

$$n_{i,ped}=n_{e,ped} \times( (z_k+1)-Z_{eff,ped})/z_k = n_{e,ped} \times( (8-Z_{eff,ped})/ z_k$$

The pedestal $Z_{eff,ped,}$ would have to increase from the unseeded level of ~1.3 (assuming a flat $Z_{eff}$ profile) to ~4 in order to achieve the required reduction in $n_{i,ped}$. With this in mind, we can now turn our attention to the rest of the profile. In the experiment, $T(r=0)/T_{ped}$~4.5 and this ratio can be assumed to stay constant as we have seen that no change in electron temperature peaking is observed in the experiment, see Fig. 8. The $Z_{eff}$ profile necessary to explain the increase in stored energy compatible with the increase in line-integrated $Z_{eff}$ measurement from 1.3 to 1.4 and $Z^{ped}{}_{eff}$~4. Such a profile is shown in Fig. 16 together with the profile of the stored energy. It can be seen that the $Z_{eff}$ profile would have to be very hollow as shown in Fig 16, to be compatible with a pedestal temperature increase of a factor of 1.4, with a central $Z_{eff}$ of 1.4 and a 1.4 increase in total stored energy. Although not measured at present in the JET-ILW, the $Z_{eff}$ profile was measured in the JET-C discharges and is slightly hollow as shown in Fig. 16. As the density and temperature gradients have not significantly changed for the considered discharges between JET-ILW and JET-C, in the region r/a~0.4-0.8, it is not expected for the impurity convection and diffusion coefficients to be much changed and as a result the peaking of $Z_{eff}$ is expected to be similar for JET-ILW and JET-C discharges.

Furthermore, if the dilution effect was the explanation for the increase in stored energy it should be independent of the plasma configuration. But we have seen that a similar rise in $Z_{eff}$ is measured in low and high $\delta$ configuration but such an increased in total pressure has not been observed. It is then highly unlikely that the increase in confinement observed in high triangularity configuration with N-seeding is related to a dilution effect. Further investigations are considering a change of pedestal stability which is outside the scope of this paper.

## 8. Conclusion

It was expected that the change from CFC to the ILW at JET would limit the operational window to higher fuelling levels both for low- and high-shape ELMy H-mode discharges, in order to keep the W contamination and induced radiation low in the main plasma. However, it was not anticipated that the new plasma-facing components would reduce the pedestal confinement of high-shape ELMy H-mode in comparison to the JET-C counterpart. The injection of an extrinsic impurity, $N_2$, in high-$\delta$ ELMy H-mode was shown to recover the pedestal confinement. This paper suggests that C content could have been a hidden parameter in the JET-C confinement behaviour of high shaped plasma and that $N_2$ partially recovers this effect in JET-ILW. The exact mechanism has not been identified so far. It was shown that once the pedestal pressure of typical discharges in JET-C is mostly recovered, the ELM governs the averaged W erosion yield, preventing stationary conditions to be achieved so far. Operation at high H-factor and high density is paramount for JET-ILW and ITER. The next step for the development of the JET-ILW compatible scenarios will be to integrate techniques to control plasma stationarity to open-up the operational space to long pulse and high current operation.

*This work, supported by the European Communities under the contract of Association between EURATOM and CCFE, was carried out within the framework of the European Fusion Development Agreement. The views and opinions expressed herein do not necessarily reflect those of the European Commission. This work was also part-funded by the RCUK Energy Programme under grant EP/I501045*

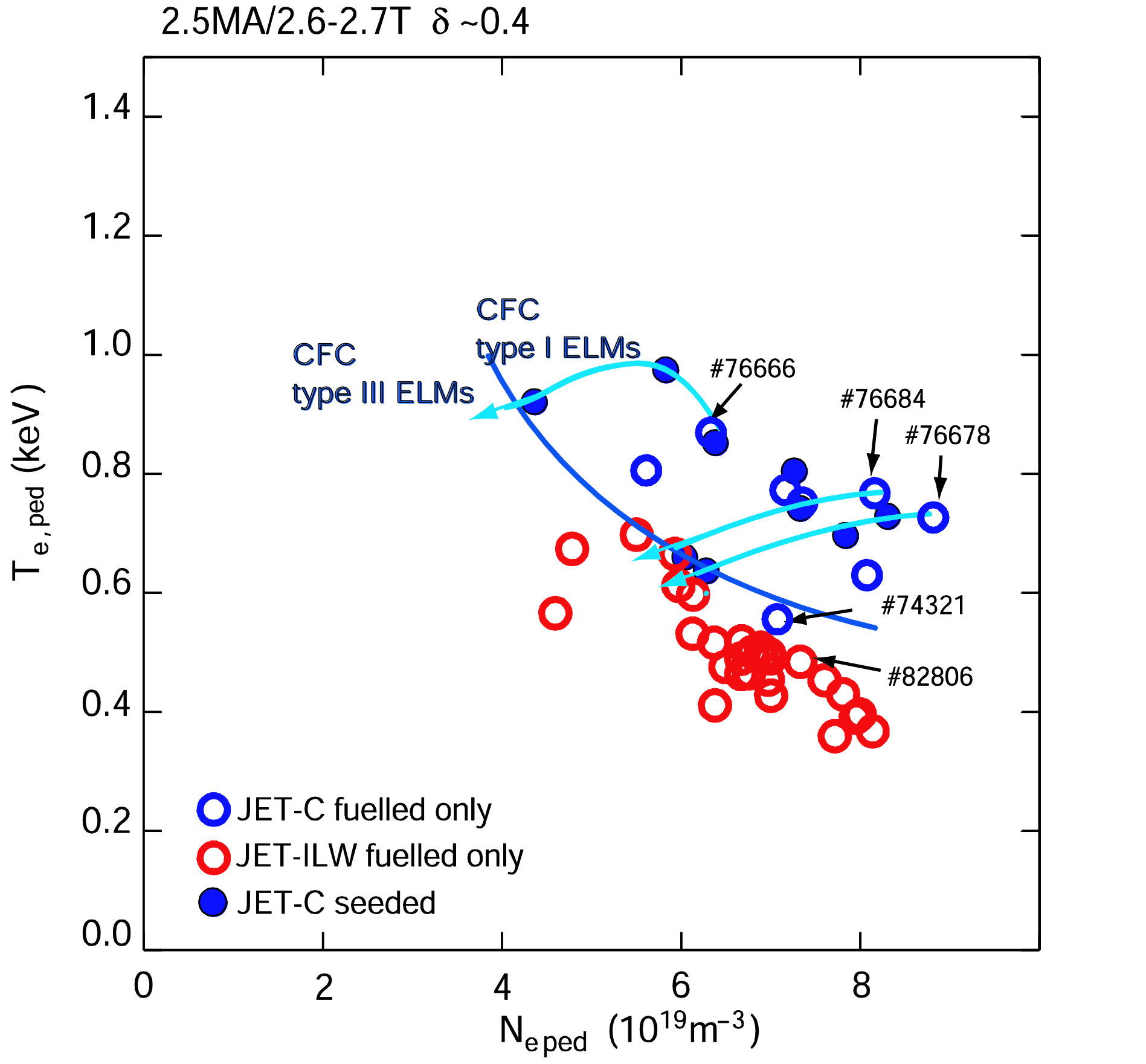

**Figure 1 Pedestal ne-Te diagram for high δ ELMy H-mode discharges for JET-C and JET-ILW discharges with 2.5MA and 2-6-2.7T:** Each symbol correspond to a discharge. JET-C discharges are in blue symbols, open if only fuelled and filled with the addition of $N_2$ seeding. The dark blue line indicates the boundary in ne-Te between type-I and type-III ELMy H-mode obtained experimentally as explained in [6]. Discharges with increasing of $D_2$-fuelling rate of 0.3, 1.8, 2,7, 6x$10^{22}$ el/s are indicated with respectively, #76666, #76684, #76678 and #74321. The light blue arrows indicate the trajectory of pedestal parameters of discharges at same $D_2$-fuelling and increased $N_2$-seeding rate from JET-C discharges [6].

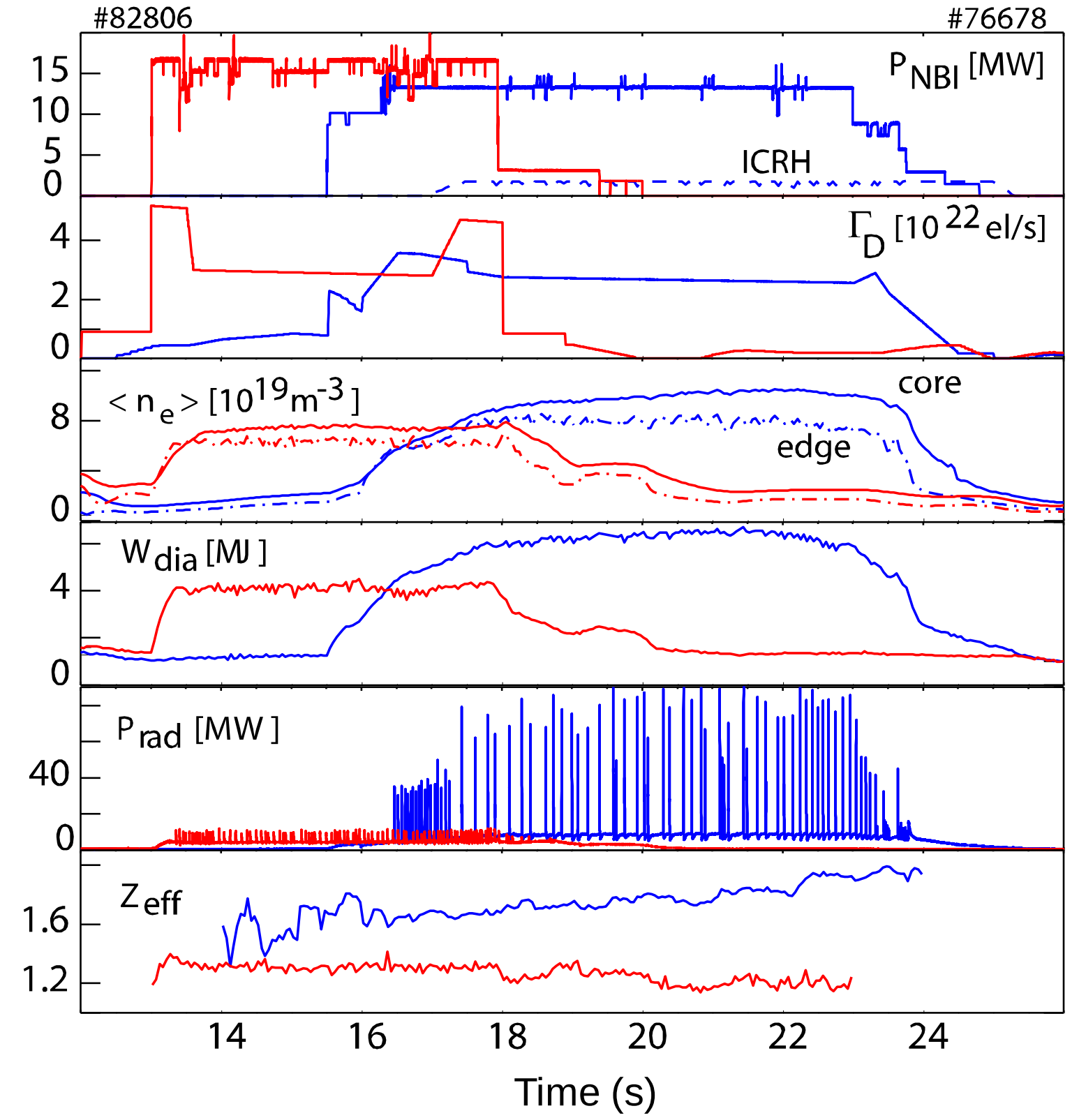

**Figure 2: Time traces of unseeded ELMy H-mode discharges in JET-C wall #76678 (in blue), and JET-ILW #82806 (in red):** (from top to bottom) NBI and ICRH (dashed) heating power, $D_2$-fuelling waveform, line-integrated core and edge (dashed) electron density measured by interferometer, diamagnetic stored energy, total radiated power and $Z_{eff}$.

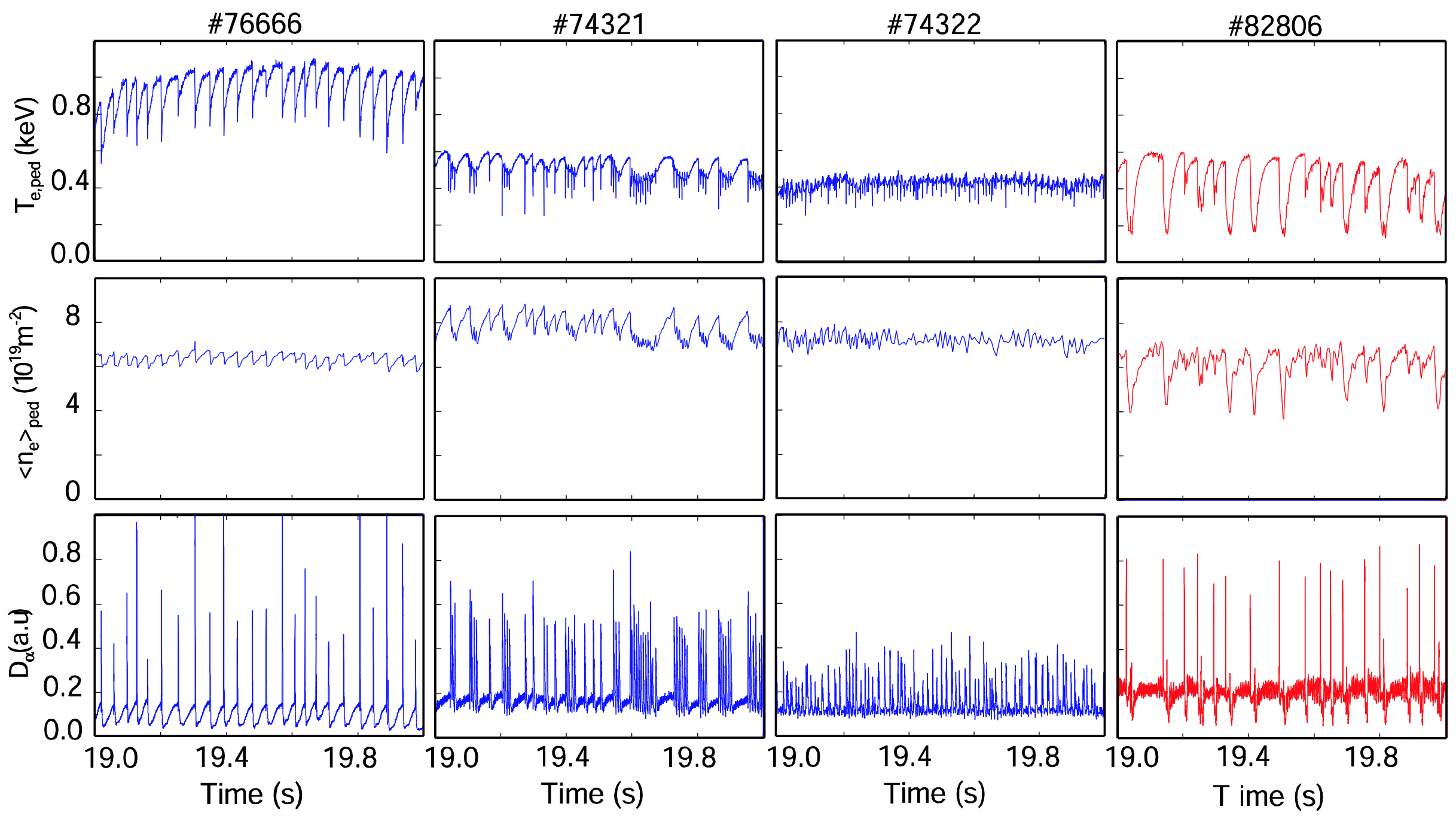

**Figure 3: Time trace of pedestal temperature, edge line-integrated density and Dα intensity**: Discharges in JET-C (in blue) are shown with typical type-I ELM (#76666), in between type-I and type-III ELM regime (#74321) and in type-III ELM regime (#74322, $T_{e,ped}$~0.55keV and $n_{e,ped}$~7.1x$10^{19}$$m^{-3}$). JET-ILW discharge #82806 (in red) is showing slow ELM collapse of electron temperature and density. The discharges electron and temperature pedestal as determined by HRTS are shown in Fig. 1 with the exception of discharge #74322.

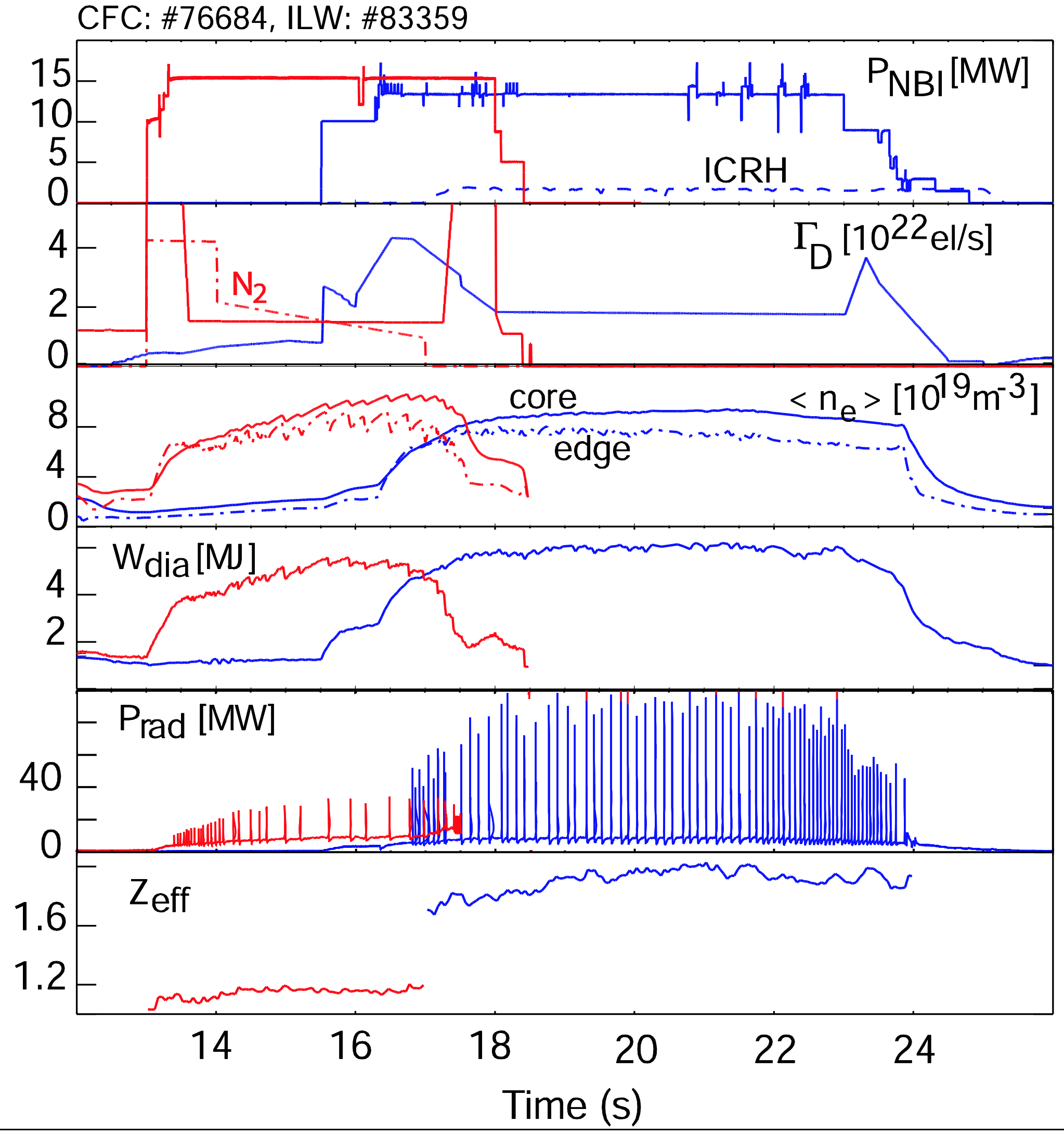

**Figure 4: Time traces of unseeded ELMy H-mode discharges in JET-C wall #76684 (in blue), and seeded JET-ILW #83359 (in red):** (from top to bottom) NBI and ICRH (dashed) heating power, $D_2$-fuelling and $N_2$-seeding (dashed) waveforms, line-integrated core and edge (dashed) electron density measured by interferometer, diamagnetic stored energy, total radiated power (without disruption phase for #83359) and $Z_{eff}$.

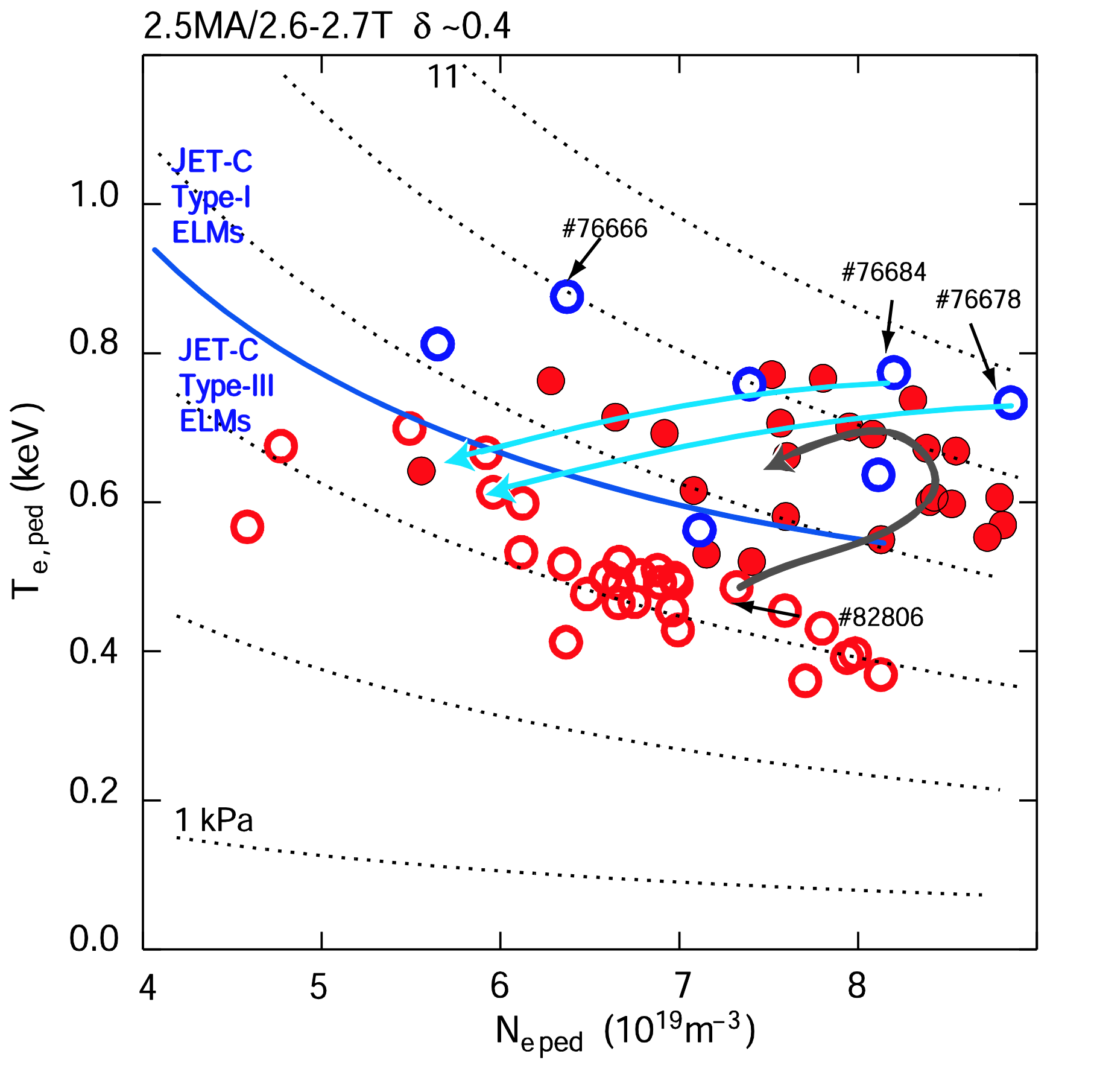

**Figure 5: Pedestal ne-Te diagram for high δ ELMy H-mode discharges for fuelled only JET-C and fuelled and $N_2$-seeded JET-ILW discharges with 2.5MA and 2-6-2.7T:** Each symbol corresponds to a discharge. JET-C discharges are in blue symbols. For clarity the $N_2$-seeded JET-C discharges have been removed and the trajectory in ne-Te diagram as $N_2$-seeded is raised in successive discharges is shown in light blue for the two highest fuelling shown in Fig.3. JET-ILW discharges are shown in red, open symbol for fuelled only discharges and filled symbol with $N_2$-seeding. The grey arrow is indicative of trajectory in pedestal density and temperature of discharges with increasing $N_2$-seeding rate from 0 to 3.7x10$^{22}$ el/s at a $D_2$-fuelling rate of 2.8x10$^{22}$ el/s.

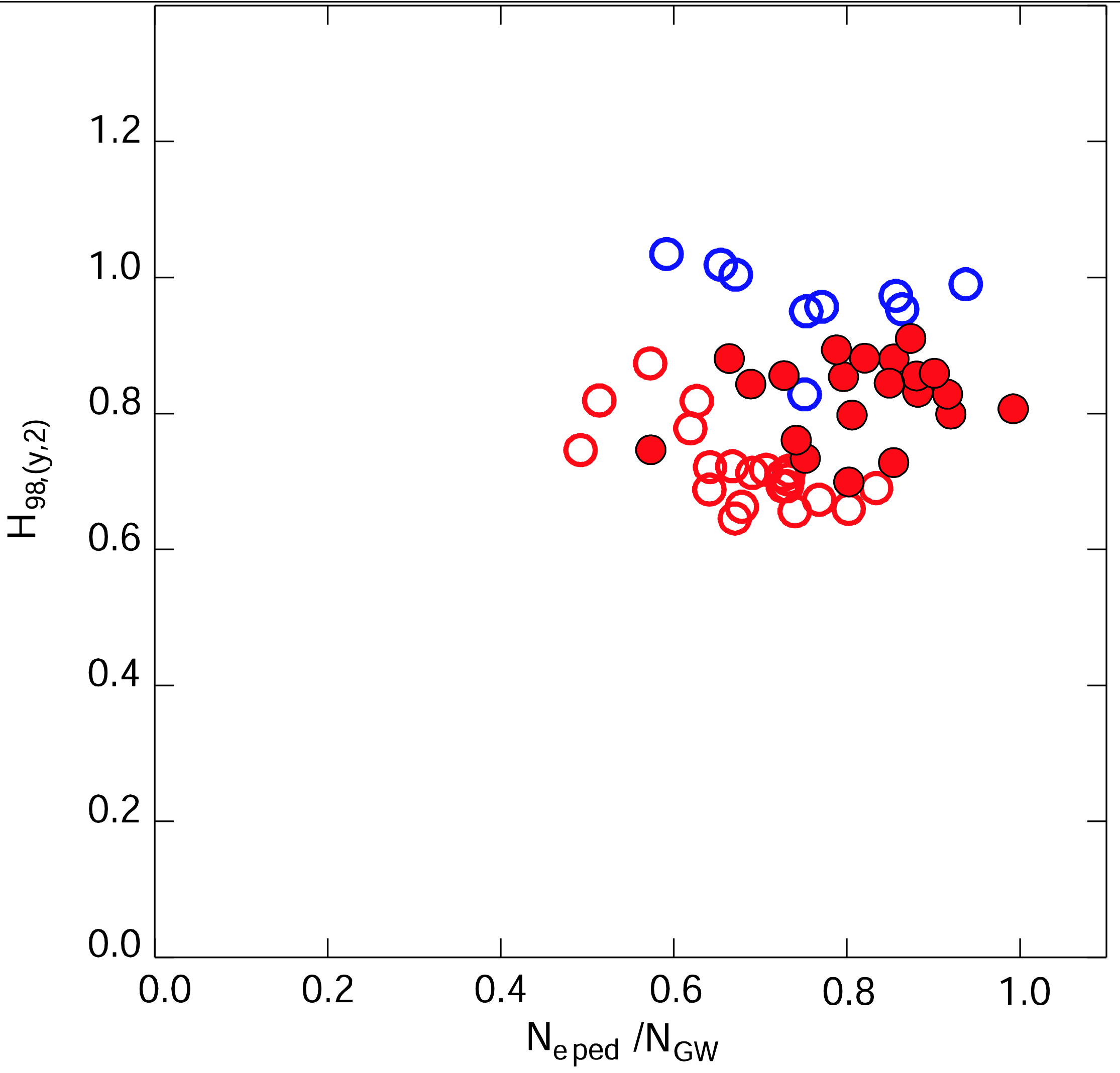

**Figure 6: Normalised energy confinement $H_{98(y,2)}$ versus pedestal density normalized to the Greenwald density**: Symbols and colors same as in Fig. 7.

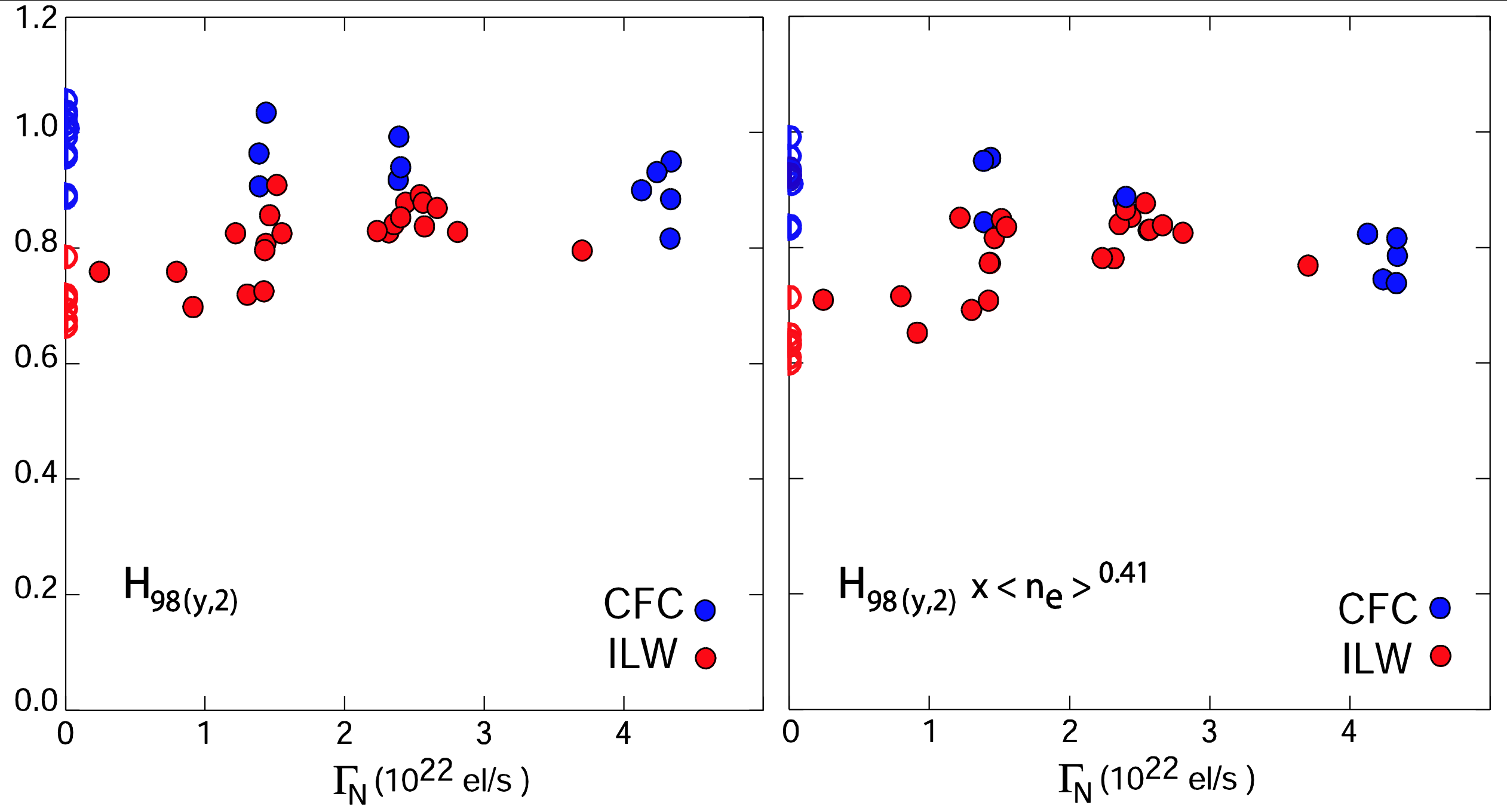

**Figure 7: Dependence of normalized confinement with $N_2$-seeding rate:** {on left} with the $H_{98(y,2)}$ density scaling and (on right) with the density scaling removed. Symbols and colors same as in Fig. 7. Filled blue symbols correspond to N2-seeded discharges in JET-C.

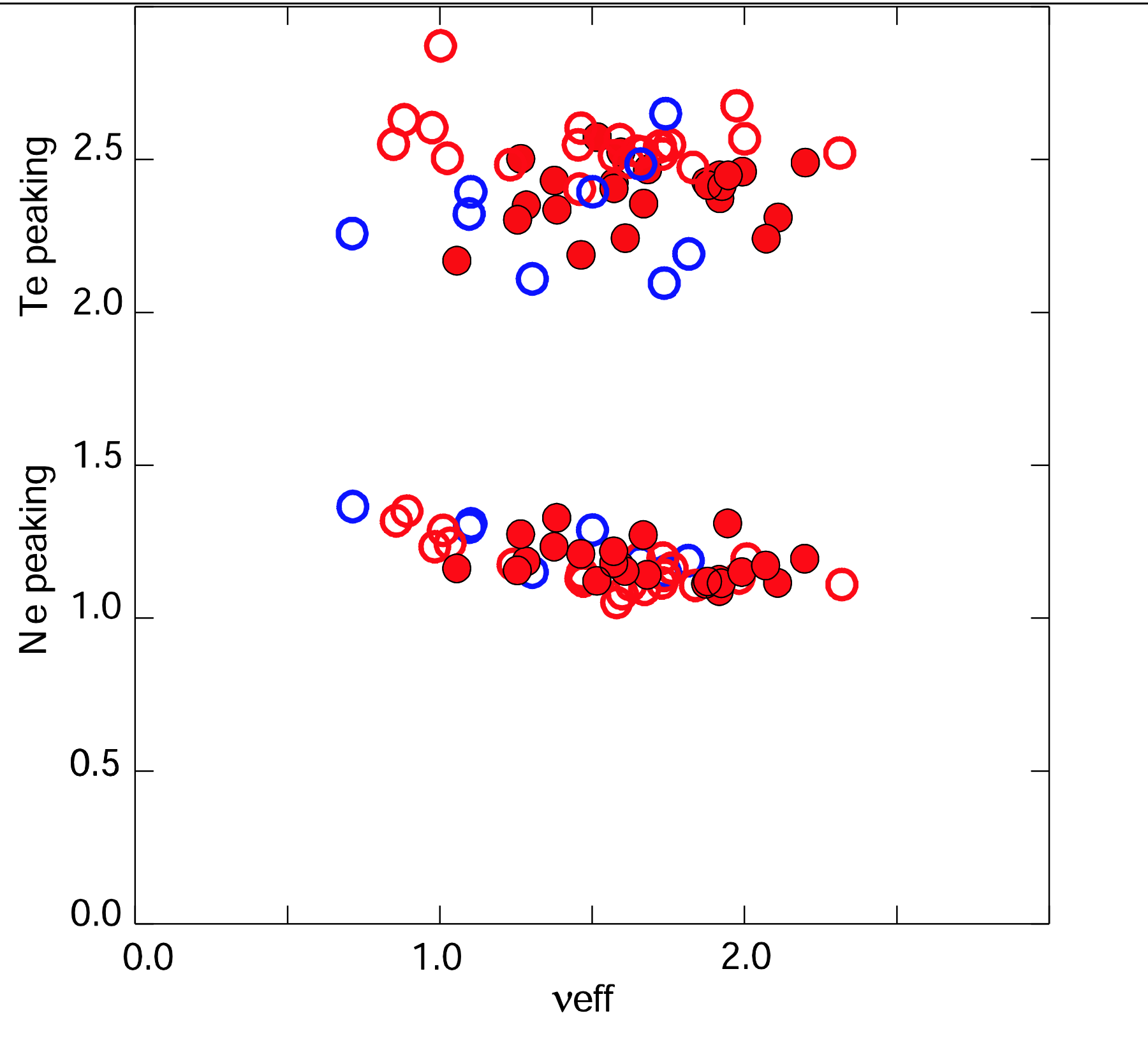

**Figure 8: Dependence of temperature peaking and density peaking with $\nu_{eff}$:** Symbols and colors same as in Fig. 7.The peaking is calculated taking the average quantity at r/a ~0.4 over its value at r./a~0.8. $\nu_{eff}$ is defined as the ratio of the electron collision frequency to the curvature drift frequency and is evaluated by the simple relation $\nu_{eff}$=0.1.R.$Z_{eff}$.$n_e$.$T_e^{2}$ with the major radius R is in m, the local density $n_e$ is in $10^{19}m^{-3}$ and the electron temperature $T_e$ is in keV.

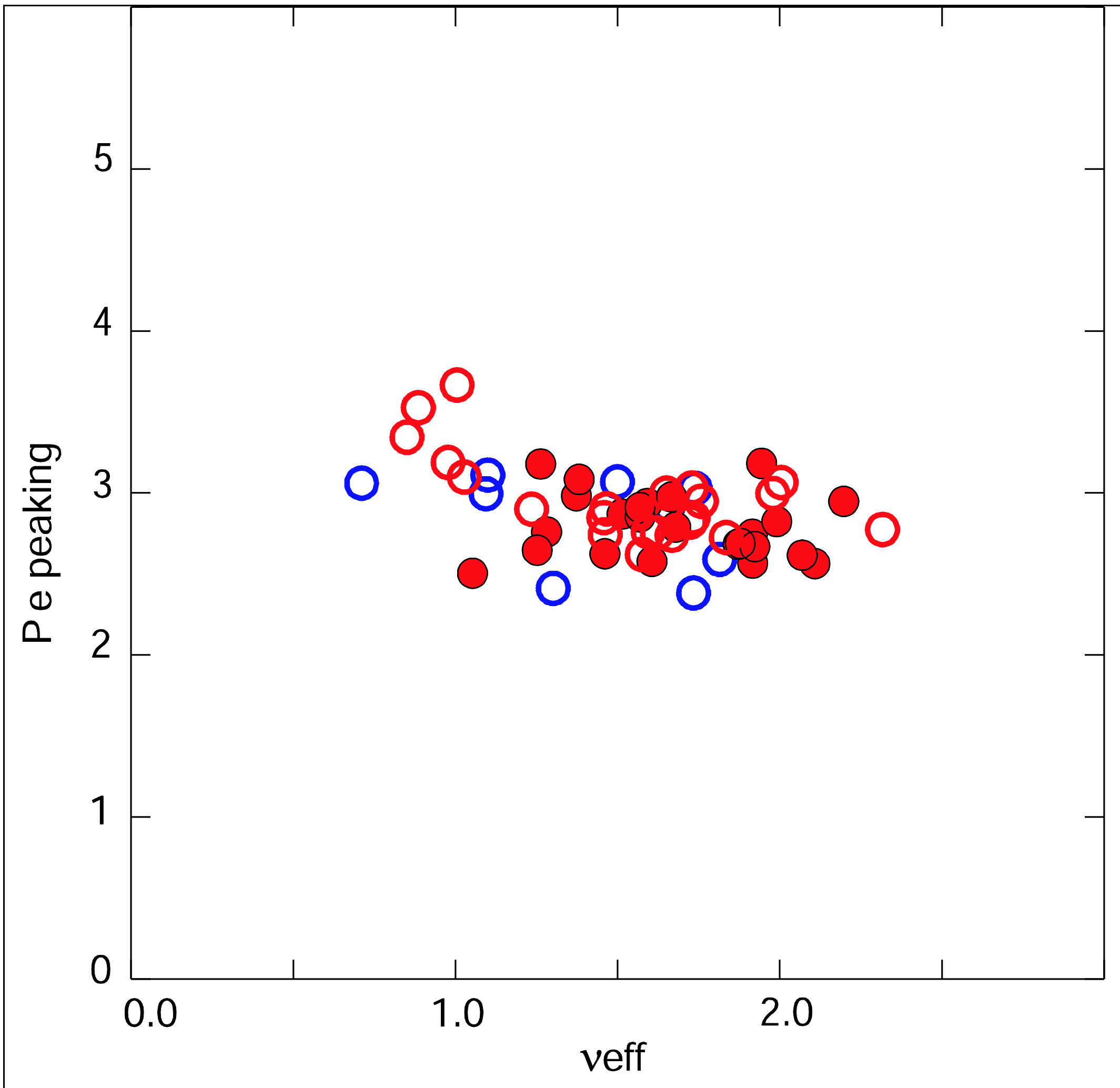

**Figure 9: Dependence of the electron pressure peaking with $\nu_{eff}$:** Symbols and colors same as in Fig. 7.

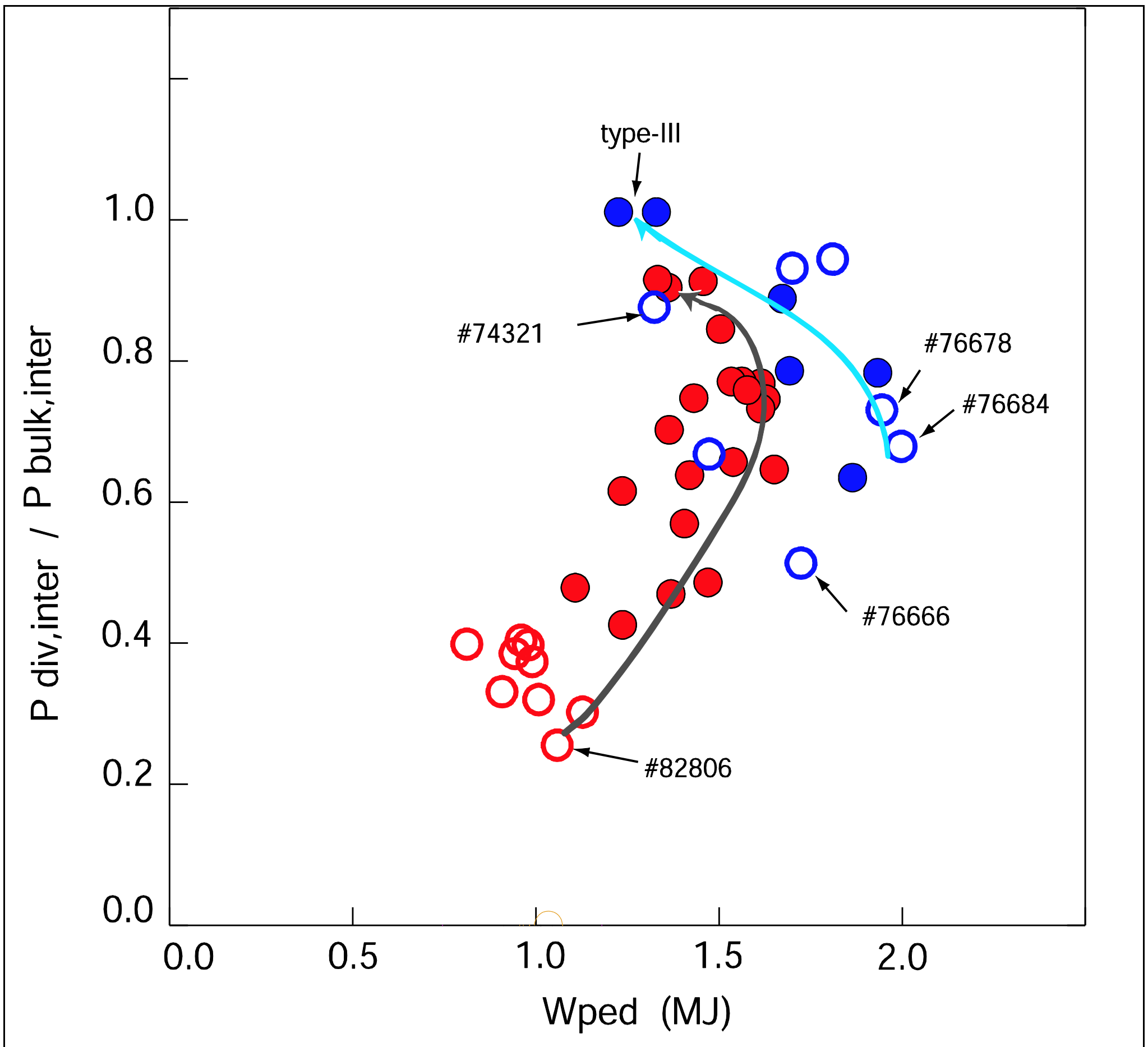

**Figure 10: Ratio of inter-ELM phase divertor radiation over main plasma radiation versus pedestal stored energy:** for JET-ILW (in blue) and JET-C (in red), with fuelling only (in open circle) and with additional seeding (in full circle). The light blue and grey arrow same as in Fig. 5. $W_{ped}$ is the pedestal stored energy and is calculated from density and temperature profiles, averaged over the time window of interest. *W*ped is equal to $3/2k\mathrm{B}(n_e \cdot T_e + n_i \cdot T_i) \cdot V_{plasma}$ with electron density $n_e$, electron temperature $T_e$, main ion density $n_i$ and ion temperature *T*i evaluated at the pedestal top and $V_{plasma}$ the total plasma volume.

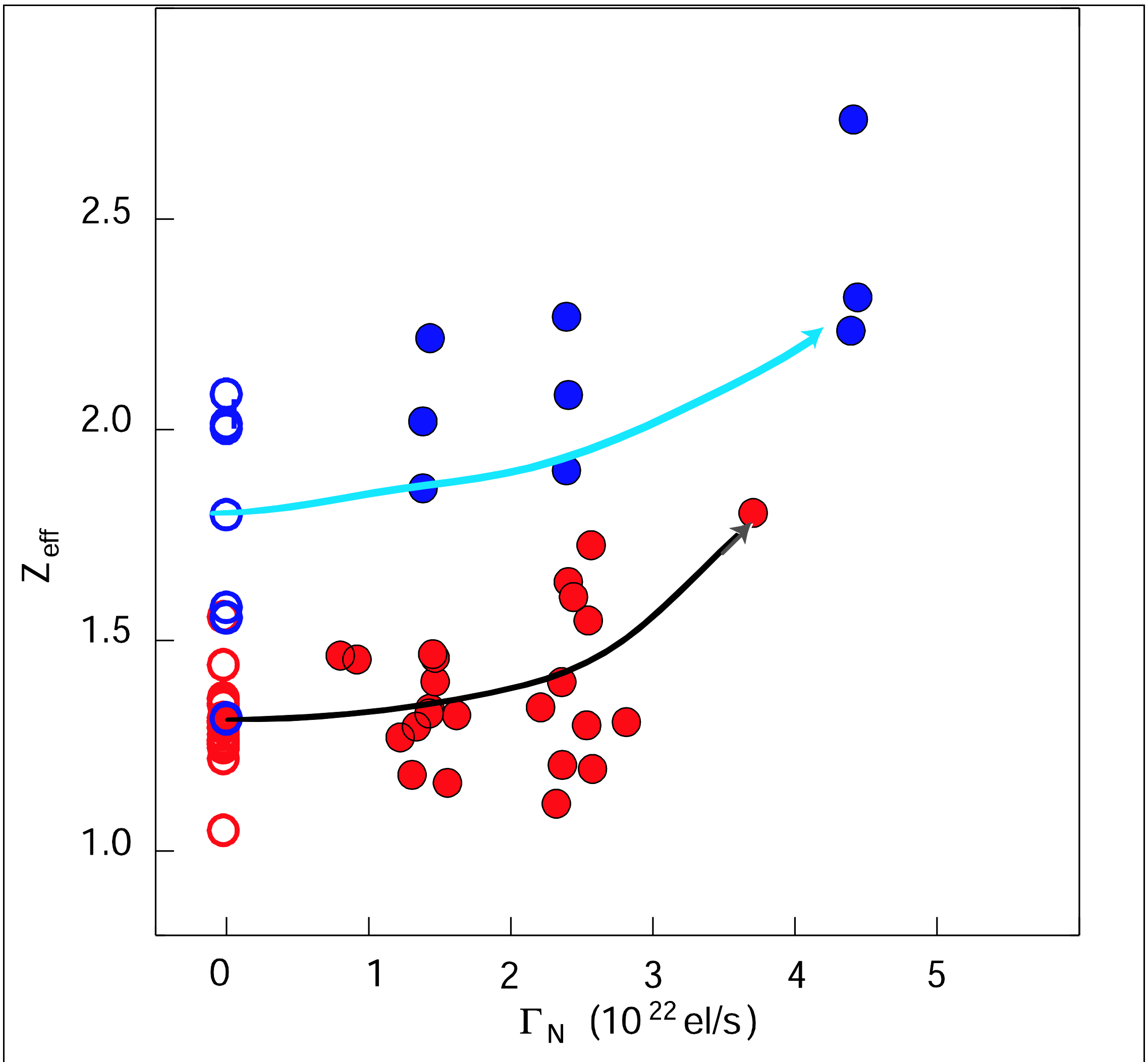

**Figure 11: Dependence of line-average $Z_{eff}$ versus $N_2$-seeding rate**: The $Z_{eff}$ is measurement from line-integrated bremsstrahlung radiation. Symbols and colors same as Fig. 12. Arrows follow the same trajectory as in Fig. 10.

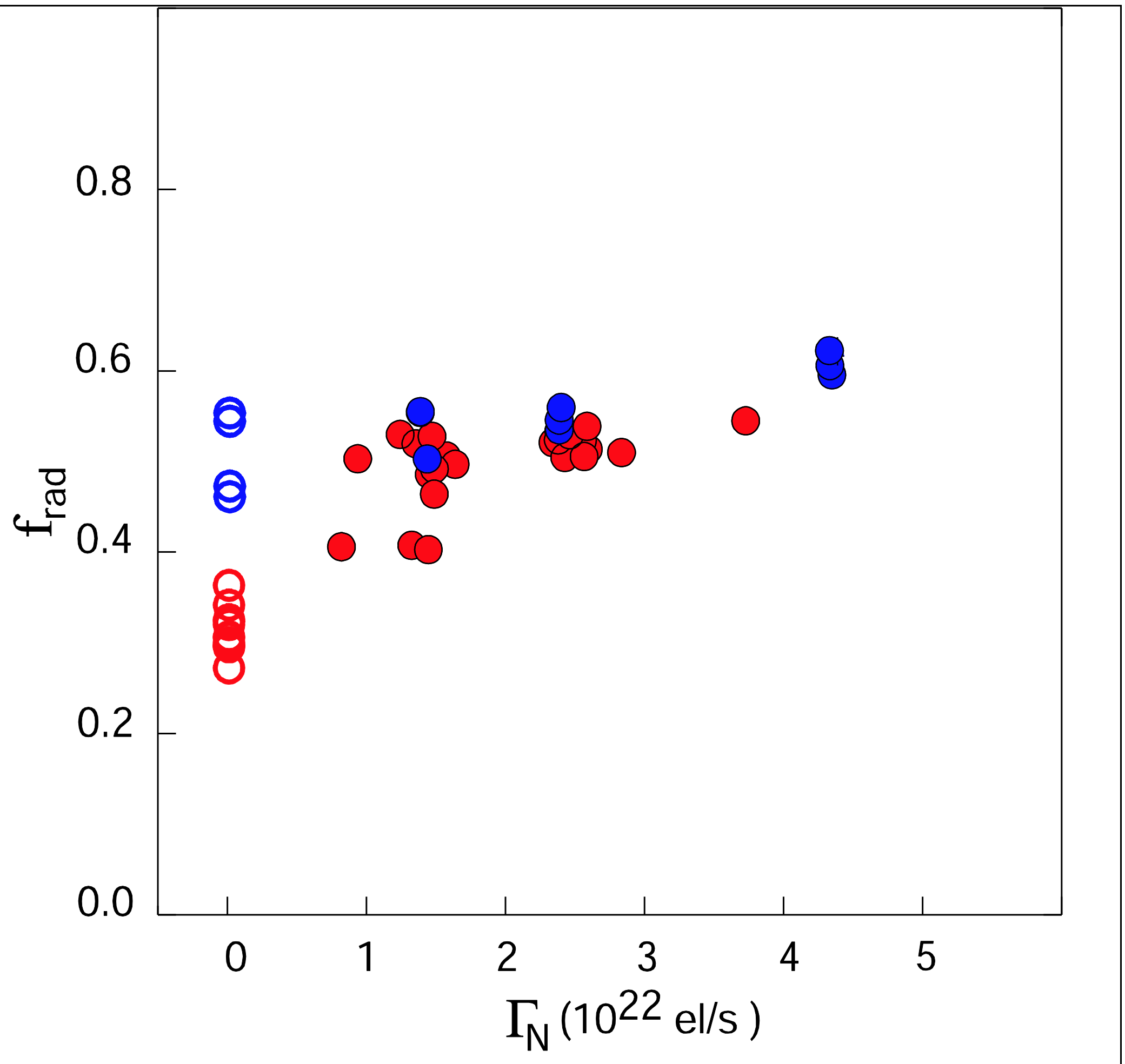

**Figure 12: Total radiative fraction versus N2-seeding rate:** Symbols and colors same as Fig. 12.

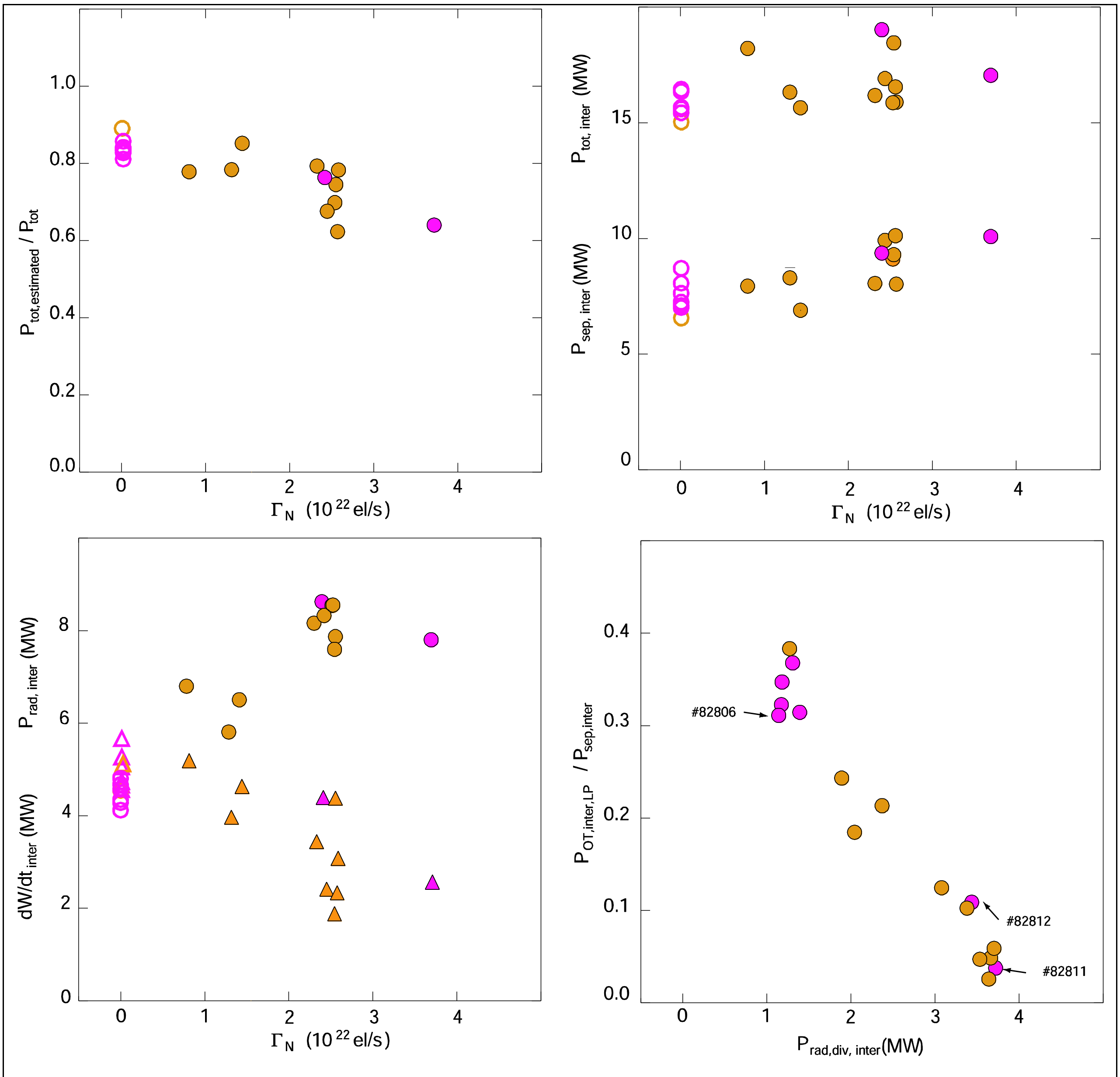

**Figure 13:** From left to right, Top row, Ratio of estimated loss power over total input power, total input power and power flowing through the separatrix, $P_{sep,inter}$. Bottom row, total radiated power and the stored energy build-up rate, the ratio of power measured by LP at outer target over $P_{sep,inter}$. All graphs show inter-ELM averaged quantities and are versus the $N_2$-seeding rate. Only JET-ILW discharges are shown here. Pink discharges correspond to discharges with D2-fuelling rate between 2.6-3.0x$10^{22}$ el/s and orange one between 0.8-2.0x$10^{22}$ el/s. Triangle are indicating the stored energy build-up rate $dW^{inter}/dt$

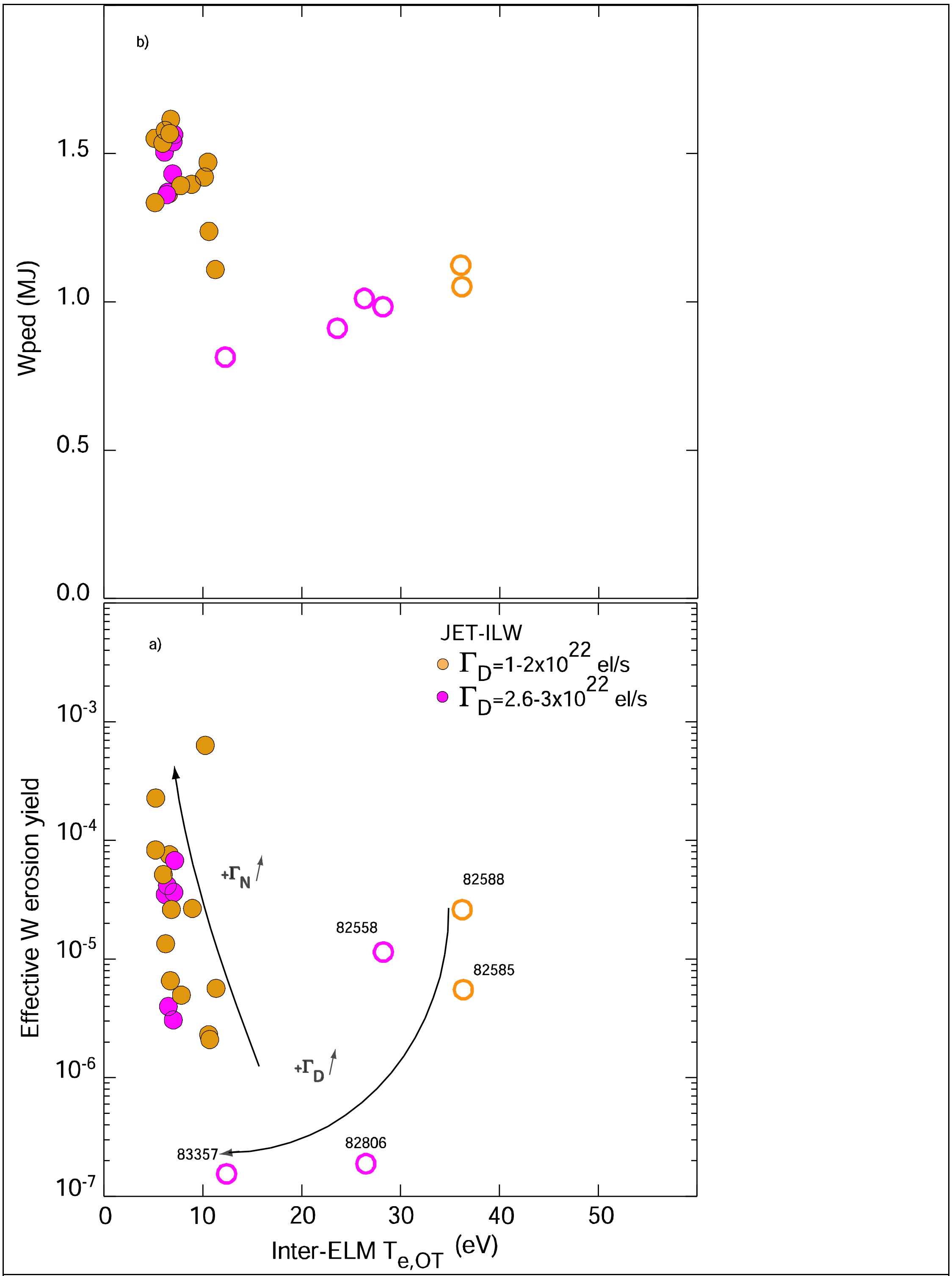

**Figure 14: Effective W erosion yield a) and average pedestal stored energy versus target electron temperature in inter-ELM period b).**

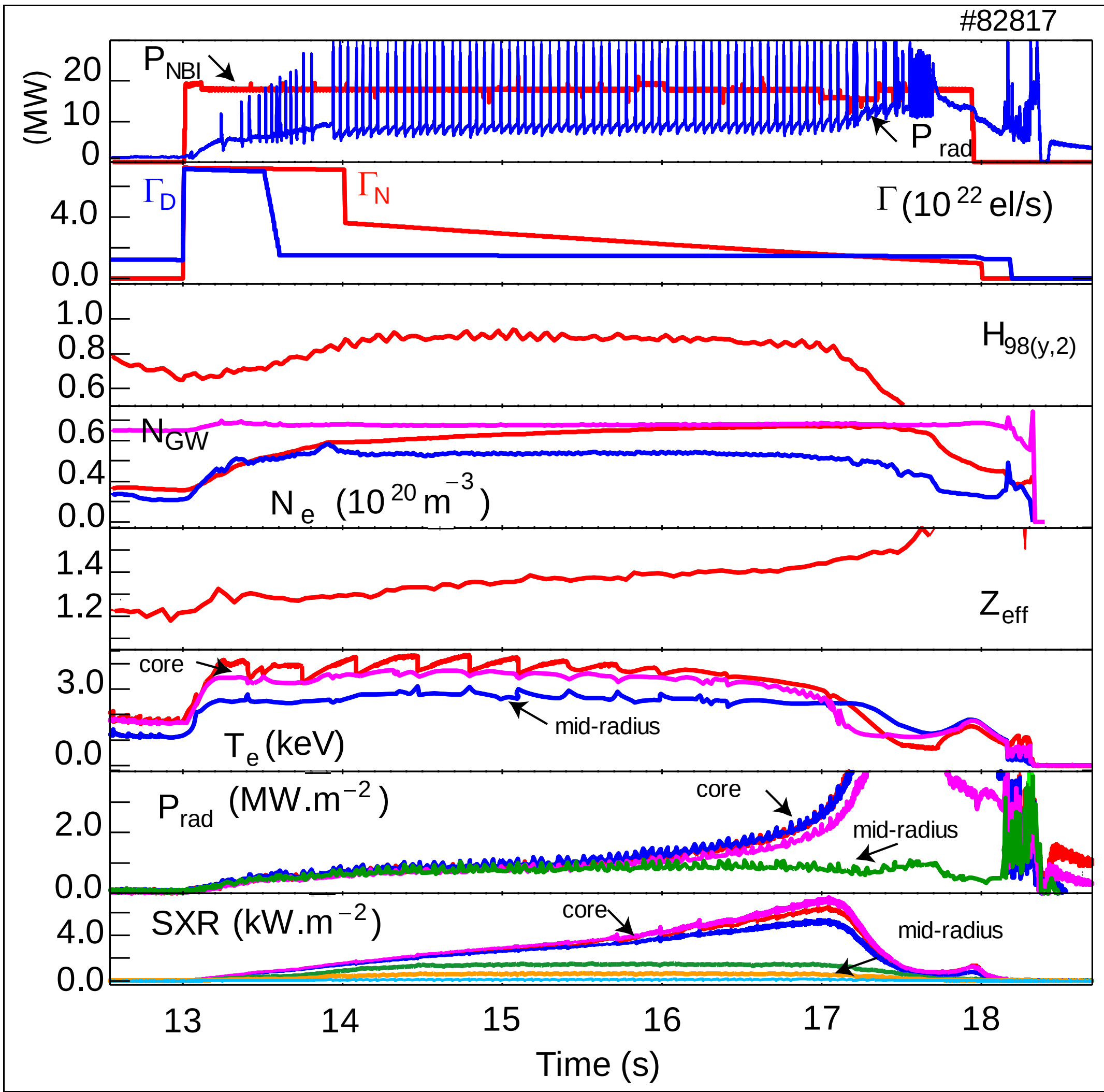

**Figure 15: Time traces of $N_2$-seeded ELMy H-mode discharges in JET-ILW wall #82817** (from top to bottom) NBI heating and total radiated power, $D_2$ and $N_2$ waveforms, normalized confinement factor $H_{98}$, core (red) edge (in blue) and Greenwald electron density (in pink), line-integrated $Z_{eff}$, electron temperature from core (in red and pink) and mid-radius (in blue), radiated power from core (in blue and pink) and mid-radius (in green) channel, and finally Soft x-ray radiated power.

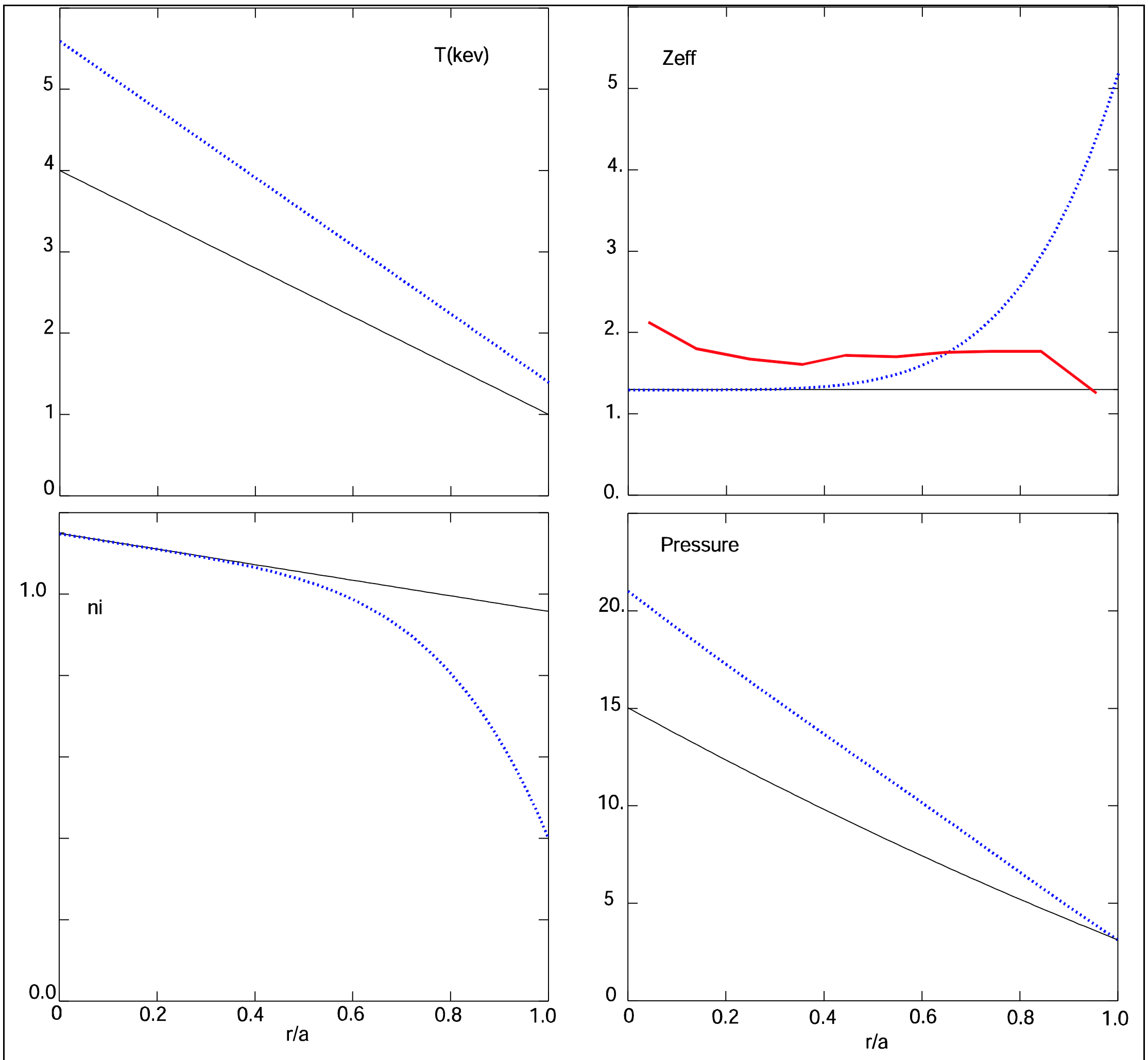

**Figure 16: Schematic of increase of pedestal temperature due to a dilution effect compatible with increase in pedestal temperature, Zeff and total stored energy seen in the experiment:** in black the example profiles before and in blue after $N_2$ seeding. Zeff profile in red is typical of those measured in JET-C seeded studies.